\documentstyle[aps,epsfig]{revtex} \def\narrowtext{} 
\begin{document} 
\newcommand{\dprime}{{\prime\prime}}
\newcommand{\be}{\begin{equation}}
\newcommand{\den}{\overline{n}} 
\newcommand{\ee}{\end{equation}}
\newcommand{\bea}{\begin{eqnarray}} 
\newcommand{\eea}{\end{eqnarray}}
\newcommand{\nn}{\nonumber} 
\newcommand{\vk}{{\bf k}}
\newcommand{\vE}{{\bf E}}
\newcommand{\vj}{{\bf j}}
\newcommand{\vs}{{\bf v}_s}
\newcommand{\vn}{{\bf v}_n}
\newcommand{\vv}{{\bf v}} 
\newcommand{\la}{\langle}
\newcommand{\ra}{\rangle} 
\newcommand{\ph}{\phi} 
\newcommand{\dg}{\dagger}
\renewcommand{\vr}{{\bf{r}}} 
\newcommand{\vq}{{\bf{q}}}
\newcommand{\vQ}{{\bf{Q}}} 
\newcommand{\hj}{\hat{\alpha}}
\newcommand{\hx}{\hat{\bf x}} 
\newcommand{\hy}{\hat{\bf y}}
\newcommand{\hz}{\hat{\bf z}}
\newcommand{\vS}{{\bf S}} 
\newcommand{\cV}{{\cal U}}
\newcommand{\cD}{{\cal D}} 
\newcommand{\tnh}{{\rm tanh}}
\newcommand{\sh}{{\rm sech}} 
\newcommand{\vR}{{\bf R}}
\newcommand{\crx}{c^\dg(\vr)c(\vr+\hx)}
\newcommand{\crkubox}{c^\dg(\vr)c(\vr+\hat{x})}
\newcommand{\pll}{\parallel} 
\newcommand{\crj}{c^\dg(\vr)c(\vr+\hj)}
\newcommand{\crmj}{c^\dg(\vr)c(\vr - \hj)}
\newcommand{\sumall}{\sum_{\vr}} 
\newcommand{\sumx}{\sum_{r_1}}
\newcommand{\nabj}{\nabla_\alpha \theta(\vr)} 
\newcommand{\nabx}{\nabla_1\theta(\vr)} 
\newcommand{\sumy}{\sum_{r_2,\ldots,r_d}}
\newcommand{\krj}{K(\vr,\vr+\hj)} 
\newcommand{\sigr}{|\psi_0\rangle}
\newcommand{\sigl}{\langle\psi_0 |}
\newcommand{\sier}{|\psi_{\Phi}\rangle}
\newcommand{\siel}{\langle\psi_{\Phi}|}
\newcommand{\sumrj}{\sum_{\vr,\alpha=1\ldots d}}
\newcommand{\krw}{K(\vr,\vr+\hx)} 
\newcommand{\Dtheta}{\Delta\theta}
\newcommand{\rhonew}{\hat{\rho}(\Phi)}
\newcommand{\rhoold}{\hat{\rho_0}(\Phi)} 
\newcommand{\dt}{\delta\tau}
\newcommand{\cP}{{\cal P}} 
\newcommand{\cS}{{\cal S}}
\newcommand{\vm}{{\bf m}} 
\newcommand{\hnr}{\hat{n}({\vr})}
\newcommand{\hnm}{\hat{n}({\vm})} 
\newcommand{\del}{\hat{\delta}}
\newcommand{\upa}{\uparrow} 
\newcommand{\dna}{\downarrow}
\draft

\title{
Phase fluctuations, dissipation and superfluid stiffness 
in d-wave superconductors
}
\author {L. Benfatto$^{1}$, S. Caprara$^{1}$, C. Castellani$^{1}$, 
A. Paramekanti$^{2}$, M. Randeria$^{2}$
       }
\address{
         $^{1}$ Universit\`a di Roma "La Sapienza"  and 
	  \\ Istituto Nazionale per la Fisica della Materia, Unit\`a di 
	  Roma 1, Piazzale Aldo Moro, 5, 00185 Roma, Italy \\
         $^{2}$ Tata Institute of Fundamental Research, Mumbai 400 005, India \\
         }

\address{%
\begin{minipage}[t]{6.0in}
\begin{abstract}
We study the effect of dissipation on quantum and thermal phase fluctuations 
in $d$-wave superconductors. Dissipation, arising from a nonzero 
low frequency optical conductivity which has been measured in experiments
below $T_c$, has two effects: 
(1) a reduction of zero point phase fluctuations, and
(2) a reduction of the temperature at which one crosses over to 
classical thermal fluctuations.
For parameter values relevant to the cuprates, we show that 
the crossover temperature is still too large for classical phase fluctuations 
to play a significant role at low temperature. Quasiparticles are thus
crucial in determining the linear temperature dependence of the in-plane 
superfluid stiffness. Thermal phase fluctuations become important at higher 
temperatures and play a role near $T_c$.
\typeout{polish abstract}
\end{abstract}
\end{minipage}}

\maketitle
\narrowtext

\noindent

\section{introduction} 

There is considerable experimental evidence 
for a linear temperature dependence of the superfluid density at
low temperatures in high-$T_c$ superconducting cuprates, i.e.
\be
\rho_s (T)= \rho_s (0) - \alpha T
\label{eq1}
\ee
where $\alpha$ is a weakly doping-dependent constant 
\cite{hardy93,panagopoulos98}.
However, there is still some controversy regarding
the low-energy excitations responsible
for this thermal suppression of $\rho_s$. The simplest explanation
is in terms of quasiparticle excitations near the
$d$-wave nodes \cite{hirschfeld93,palee97,millis98,mesot99}.
An alternative explanation is in terms
of thermal fluctuations of the phase of the order parameter 
\cite{roddick95,emery95,carlson99} or other collective modes 
\cite{levin98}. 

The existence of well-defined
quasiparticles in the superconducting state of cuprates is supported both by
transport \cite{bonn98,ong99,chiao99} and ARPES \cite{kaminski99} experiments
(even though there are some studies questioning their Fermi liquid description
\cite{valla99,corson00}).  
However, the contributions of phase fluctuations to low temperature properties
could still be important, especially in the underdoped regime where the 
superfluid density $\rho_s$ becomes vanishingly small as the Mott insulator 
is approached.

The study of phase fluctuations raises several important issues:
(1) The form of the phase-only action 
for layered $d$-wave superconductors (SC's) taking 
into account the long-range Coulomb interaction; 
(2) The crossover between quantum and classical regimes of phase fluctuations.
These questions were studied in detail in ref.~\cite{arun00} which, however,
did not discuss the role of dissipation.
In this paper, we focus on dissipative effects and how they affect 
the form of the phase-only action and the quantum-to-classical crossover.

There are several reasons to believe that low energy dissipation is important 
in the high-$T_c$ cuprates. Theoretically,
weak disorder within a self-consistent T-matrix calculation
leads to a nonzero ``universal'' low frequency quasiparticle conductivity 
\cite{palee93} in $d$-wave SC's. 
Experiments have also measured a nonzero low frequency conductivity 
\cite{sflee96,corson00}, much larger than the ``universal'' value.
While there is no consensus on the origin of this large conductivity,
one would definitely expect this dissipation to affect the phase fluctuations
in the system, as first emphasized by Emery and Kivelson (EK) \cite{emery95}.
Our formalism and results, however, differ from those of EK as discussed
in detail in the paper.

We summarize our main results below:

(1) We derive the Gaussian effective action for phase fluctuations 
in the presence of dissipation using a functional integral approach and 
integrating out fermionic degrees of freedom. 
While our effective action is derived
microscopically by looking at fluctuations around a BCS mean field solution, 
we make contact with experiment by using parameter values
relevant to the high $T_c$ SCs. 
We believe this phenomenological approach of using the derived {\em form} 
for the action, with {\em coefficients} taken from experiment, is
valid for the SC state of the high $T_c$ materials, at least for $T \ll
T_c$, when quasiparticles are well defined.
In addition, we also present in an Appendix, a hydrodynamic derivation for the 
phase mode based on a two-fluid model, which serves as a check on the
microscopic derivation. 

(2) A dissipative quantum $X Y$ model is obtained by coarse-graining 
the Gaussian action to the scale of the coherence length and analyzed 
within a self-consistent harmonic approximation.

(3) We find that the magnitude of quantum fluctuations at $T=0$ is
reduced by the presence of dissipation. 

(4) Dissipative phase fluctuations alone, in the absence of quasiparticle 
excitations, are shown to lead to a $T^2$ reduction \cite{millis98} of the 
superfluid stiffness. This behavior crosses over to a classical linear $T$
reduction at a scale $T_{\rm cl}$, which decreases with increasing dissipation.

(5) Choosing parameters appropriate to the high-$T_c$ cuprates, and
overestimating the dissipation, we nevertheless find that the
crossover scale $T_{\rm cl}$ is still fairly large. Thus one cannot attribute
the low temperature linear reduction of $\rho_s(T)$ to classical
phase fluctuations. This $T$-dependence must therefore arise entirely
from quasiparticle excitations near the $d$-wave nodes.

\section{effective phase action}

We find it convenient to express the superfluid density $\rho_s$
in terms of a stiffness $D_s = \hbar^2\rho_s/m^*$, which 
in the London limit is related to the 
penetration depth $\lambda$ in a 3D bulk system through 
$1/\lambda^2=4\pi e^2 D_s/\hbar^2 c^2$. The in-plane superfluid stiffness in 
a layered system, with interlayer spacing $d_c$, is $d_c D_s$ 
with dimension of energy. Henceforth, unless explicitly displayed, we set 
$\hbar=k_{_B}=1$.

We begin with the Gaussian phase action for a 3D isotropic superconductor 
(SC)
\be
S_G[\theta] = \frac{a^3}{8 T}{\sum_{\vq,\omega_n}} \left(
\omega_n^2 \chi + D(i\omega_n) \vq^2 \right)
\theta(\vq,i\omega_n) \theta(-\vq,-i\omega_n).
\label{Gaussian}
\ee
For a derivation in the $s$-wave case see refs.~\cite{tvr89,depalo99}
and for the $d$-wave case see ref.~\cite{arun00}. We show in Appendix A 
that the above action, and its generalization to layered systems, can be 
also derived from hydrodynamic considerations within a two-fluid model.
In the above action (\ref{Gaussian}), the compressibility $\chi(\vq \to 0) 
\simeq 1/V_\vq$ where
$V_\vq$ is the Coulomb interaction and $a$ is the
lattice spacing. On continuing to real frequency $D(\omega)$ 
is the mean field stiffness, which is related to the mean-field
complex conductivity $\sigma(\omega)$ through $D(\omega)=
(-i \omega \sigma(\omega)/e^2)$. 
In arriving at the above action, we have made the implicit assumption 
that $\sigma(\vq,\omega)\approx \sigma(0,\omega)$, and ignored the 
$\vq$-dependence of the conductivity \cite{footnote.nonlocal}
for $\vq \lesssim \pi/\xi_0$. 

We use the spectral representation for $\sigma$, 
and find that 
\be
D(i\omega_n)= D_s^0 + \frac{1}{e^2}
\int_0^\infty\frac{d\omega}{\pi}\frac{2\omega_n^2}{(\omega^2+\omega^2_n)}
{\rm Re}\sigma_{reg}(\omega)
\label{spectral}
\ee
where we have used 
${\rm Re}\sigma(\omega) = \pi D^0_s e^2\delta(\omega)+{\rm Re}\sigma_{reg}(\omega)$. 
For a frequency-independent ${\rm Re}\sigma_{reg}(\omega) = \sigma_{_{DC}}$, 
this simplifies to
\be
D(i\omega_n)=D^0_s+ \frac{\sigma_{_{DC}}}{e^2} |\omega_n|.
\ee
Unless indicated otherwise, we will use this simplified form of the
conductivity below, and use
$\overline{\sigma}=\sigma_{_{DC}} d_c/(e^2/h)$ as a
dimensionless measure of the dissipation.  

It is straightforward to generalize the above results to a layered
system with an in-plane stiffness $D^0_{_\pll}$ and a c-axis stiffness 
$D^0_{_\perp}$. Further, the Coulomb interaction in a system with layer 
spacing $d_c$ gets modified to \cite{fetter74}
\be
V(\vq) = \frac{2 \pi e^2 d_c}{q_\pll\epsilon_\infty}
\left[\frac{\sinh(q_\pll d_c)}{{\rm cosh}(q_\pll d_c)-\cos(q_\perp
d_c)}\right]
\label{coulomb}
\ee
where $q_\pll,q_\perp$ are the in-plane and c-axis components of $\vq$
respectively.

To investigate the contribution of the phase fluctuations to the depletion of
superfluid density it is necessary to go beyond the Gaussian
approximation. The simplest model which allows
for such an analysis is the quantum $X Y$ model, in which the phase field
is defined on a coarse-grained lattice, with an in-plane lattice
constant of $\xi_0$ and layer spacing $d_c$. The coherence length enters as
a short distance cutoff since the mean field assumption
of a constant amplitude breaks down at shorter distances.

Following exactly the same procedure of coarse-graining 
used in ref.~\cite{arun00} (for the non-dissipative case) we now obtain
the dissipative quantum $X Y$  action:
\bea
S_{XY}[\theta] &=& \frac{1}{8 T}{\sum_{\vQ,\omega_n}}' 
\left(\frac{\omega^2_n \xi_0^2 d_c}
{\tilde{V}(\vQ)}+\frac{\overline{\sigma}}{2\pi}|\omega_n|\gamma_{_\pll}(\vQ)\right)
|\theta(\vQ,\omega_n)|^2
+ \frac{D^0_{_\pll} d_c}{4} \int_0^{1/T}~d\tau \sum_{\vR,\alpha=x,y} 
\left(1-\cos[\theta(\vR,\tau)-\theta(\vR+\alpha,\tau)] \right) \nn \\
&+& \frac{D^0_{_\perp} d_c}{4}\left(\frac{\xi_0}{d_c}\right)^2
\int_0^{1/T}~d\tau \sum_{\vR} 
\left(1-\cos[\theta(\vR,\tau)-\theta(\vR+\hz,\tau)] \right).
\label{xy1}
\eea
Here $\gamma_{_\pll}(\vQ)=(4-2 \cos Q_x-2 \cos Q_y)$ with $\vQ$ being the 
dimensionless momentum, and the scaled interaction \cite{footnote1} 
$\tilde{V}(\vQ) \equiv V(\vQ_{_\pll}/\xi_0,Q_{_\perp}/d_c)$.
While all momenta with $|\vQ_x|,|\vQ_y|,|\vQ_z| \leq \pi$ contribute in 
(\ref{xy1}) above, the
prime on the summation denotes a Matsubara frequency cutoff
discussed below (see also ref.~\cite{arun00}).

In this derivation we have promoted the gradient terms in the Gaussian action 
arising from the superfluid stiffness to the cosine form, while the 
dissipative terms have still been retained at Gaussian level. 
A more sophisticated approach would probably end up with a 
$\tau$ non-local kernel within the cosine term; 
we will however continue to work with the simplest action above.
This action (\ref{xy1}) is well known in the literature as the
resistively shunted Josephson junction (RSJJ) model and its 
phases and quantum phase transitions have 
been extensively studied \cite{chakravarty86,chakravarty88}. Here we are 
interested in the effect of dissipation on quantum phase fluctuations
and the classical crossover temperature, in the superconducting state. 

We now discuss the differences between our
action (\ref{xy1}) and that considered by Emery and Kivelson (EK)
\cite{emery95}. EK included the effects of screening 
by replacing the $V(\vQ)$ appearing in Eq.~(\ref{Gaussian})
with the screened interaction $V_s(\vQ)= V(\vQ)/\epsilon(\omega)$. 
Here $\epsilon(\omega)= 1 + 4\pi{\mathrm i} \sigma_L(\omega)/{\omega}$ 
is the dielectric function at $\vQ=0$,
and $\sigma_L(\omega)$ is the longitudinal optical conductivity. 
Considering the isotropic 3D Coulomb interaction, 
$4\pi e^2/\epsilon_\infty\vQ^2$ , the 
dynamical term in the EK analysis reduces to the form 
$\omega^2_n \epsilon_\infty \epsilon(\omega) \vQ^2/4\pi e^2$ .
This expression has been shown to be correct 
\cite{dattu98,depalo99} when the screening in the
superconductor is produced by some external degrees of freedom 
with conductivity $\sigma_L(\omega)$. An example is provided by a
coupled system consisting of a
superconductor interacting via Coulomb interactions with 
a normal metal.

However, as discussed in ref.\cite{arun00}, the EK effective action is
{\it not} obtained for a single homogeneous SC. The longitudinal conductivity
$\sigma_L$ of the SC does not explicitly appear in the expression of the 
density-density correlation function $\chi$. Instead dissipation 
appears through the transverse current-current correlation function, 
and affects the gradient term in the action (\ref{Gaussian}), so that 
$ D_{_\pll} \to D_{_\pll}-{\rm i}\omega\sigma_T(\omega) /e^2$, where
$\sigma_T$ is the transverse conductivity. 

If we would {\em assume} that physical (i.e., gauge-invariant) correlation 
functions  
appear as the coefficients in the phase action, {\em then}, using the
equality of the physical longitudinal and transverse conductivities,
it is easy to see that our action (\ref{Gaussian}) is identical to
the EK action \cite{emery95}, for the specific case of a SC with isotropic
3D Coulomb interactions. In this 3D case one could associate the $\sigma$ with
either the gradient term, as we do, or with the time derivative term,
as done by EK. The action used by EK is then formally the same as our 
action, and one could argue that dissipation should appear in the same way 
whether it is from an external bath (EK) or from internal degrees of freedom 
(our case).

The above assumption of gauge invariant coefficients is, however, not valid for a 
single homogeneous SC where the screening arises from the (low energy) internal 
degrees of 
freedom. The coefficients in the phase action are then {\it mean-field}
correlation functions; they {\em cannot}, in general, be 
gauge-invariant since the phase variable is yet to
be integrated out. It is only upon integrating out the phase variable
that one restores gauge invariance \cite{arun00}. 

\section{Quantum and Classical Phase Fluctuations}

\subsection{Variational Analysis}

We analyze the quantum $X Y$  action within the self consistent harmonic
approximation (SCHA). We believe that this is adequate to
calculate the effects of phase fluctuations at low temperatures,
where longitudinal (``spin-wave'') fluctuations dominate and 
transverse (vortex) excitations are exponentially suppressed given their 
finite core energies.
To examine the low temperature in-plane properties, we assume 
$D^0_{_\perp}=0$ in
(\ref{xy1}) since it is very small in highly anisotropic systems with 
a large $\lambda_{_\perp}$.
For parameter values appropriate to $Bi2212$, we have numerically checked 
that setting $D^0_{_\perp}=0$ does not affect our in-plane results.

The SCHA \cite{wood82,chakravarty88,roddick95} 
is carried out by replacing the above 
action by a trial harmonic theory with the
renormalized stiffness $D_{_\pll}$ chosen to minimize the 
free energy of the trial action.
This leads to 
\be
D_{_\pll} = D^0_{_\pll} \exp(- \la\delta\theta^2_{_\pll}\ra/2)
\label{scha}
\ee
where $\delta\theta_{_\pll} \equiv (\theta_{\vr,\tau}-\theta_{\vr+\alpha,
\tau})$ with $\alpha=x,y$
and the expectation value evaluated in the renormalized harmonic theory
is given by
\be
\la\delta\theta^2_{_\pll}\ra = 2 T \int_{-\pi}^{\pi}~\frac{d^3\vQ}{(2\pi)^3} 
\sum^{n_c}_{n=-n_c} \frac{\gamma_{_\pll}(\vQ)}{\omega^2_n 
\xi_0^2 d_c/\tilde{V}_\vQ + (D_{_\pll} d_c
+\frac{\overline{\sigma}}{2\pi}|\omega_n|)\gamma_{_\pll}(\vQ)}.
\label{flucn}
\ee

As mentioned earlier, the dynamical phase distortions should have a 
frequency cutoff for the simple action we have considered. In our
numerics, we use a cutoff\cite{arun00} $n_c$ corresponding to $\omega_n 
\lesssim 
\sqrt{(D_{_\pll}d_c)(2\pi e^2/\epsilon_\infty \xi_0)}$,
but we have checked that the presence of a finite $n_c$ has only a minor 
quantitative effect on the results for $\la\delta\theta^2_{_\pll}\ra$ 
in the presence of dissipation, and one may set $n_c \to \infty$ to obtain 
qualitatively correct results.

\subsection{Analytical estimates of quantum and thermal fluctuations}

We first present estimates of the magnitude of 
quantum fluctuations and the thermal crossover scale
making certain simplifying assumptions.
The in-plane quantum fluctuations are seen to be 
dominated by relatively large $Q_{_\pll}$ from phase space considerations 
and the form of the integrand in (\ref{flucn}).
In this case, we may set $\tilde{V}(\vQ\sim 1)/\xi_0^2 d_c 
\approx 2\pi e^2/\epsilon_\infty \xi_0$. Restricting
ourselves to low $T$, we ignore the Matsubara cutoffs 
and set $n_c \to \infty$. With these simplifications, we work in the 
limiting cases of small and large dissipation.
We report further analytical results in Appendix B. In particular,
we calculate the renormalization of the superfluid stiffness
for an anisotropic 3D Coulomb interaction (instead of the Coulomb interaction 
in layered systems used in the paper) which permits us to analyze the case of 
arbitrary $\overline{\sigma}$.

First recall the non-dissipative case \cite{arun00} 
where the problem involves only two energy scales: 
the Coulomb energy $(2\pi e^2/\epsilon_\infty\xi_0)$ and the
layer stiffness $D_{_\pll} d_c$. 
The quantum zero point
fluctuations of the phase are given by the dimensionless combination
$\sqrt{(2\pi e^2/\epsilon_\infty \xi_0)/D_{_\pll}d_c}$, while the 
crossover to classical fluctuations takes place at a temperature 
$T_{\rm cl} \sim \sqrt{(D_{_\pll}d_c)(2\pi e^2/\epsilon_\infty \xi_0)}$.
Taking into account the temperature dependence of the bare stiffness,
a better estimate of the crossover temperature is $T_{\rm cl} \sim T_c$.

For the dissipative case at $T=0$ we convert the 
Matsubara sum to an integral. For large $\overline{\sigma}$, ignoring 
$D_{_\pll}d_c$ in the integrand, and introducing a lower frequency 
cutoff, $2\pi D_{_\pll}d_c/\overline{\sigma}$, it is easy to show that the
the magnitude of quantum fluctuations may be estimated as $\la
\delta\theta^2_{_\pll}\ra\simeq 
\left({8\over{\overline{\sigma}}}\right)
\ln\left[\frac{\overline{\sigma}}{2\pi}
\sqrt{(2\pi e^2/\epsilon_\infty\xi_0)/ (D_{_\pll}d_c)}\right]$. 
This is similar to the result obtained
by Chakravarty {\it et al} \cite{chakravarty88} for a RSJJ model
with short range charging energies. Increasing dissipation thus leads 
to a decrease in quantum fluctuations as the system becomes more 
classical. 

To evaluate the temperature scale at which one crosses over to classical
fluctuations in the presence of dissipation, we have to consider
the temperature above which only the $n=0$ Matsubara frequency
contributes, so that phase dynamics is unimportant. For large
$\overline{\sigma}$, this may be estimated in a simple manner by setting
$(\overline{\sigma}/2\pi) |\omega_n| \gtrsim D_{_\pll} d_c$ 
with $n=1$, which ensures
that fluctuations with $n \gg 1$ would contribute very little to the 
fluctuation integral in (\ref{flucn}). This leads to
$T_{\rm cl}\gtrsim D_{_\pll}d_c/\overline{\sigma}$. A better estimate is
obtained below, which gives $T_{\rm cl} \approx 3 D_{_\pll}d_c/\overline
{\sigma}$. It is clear that the classical limit emerges as the 
limit of infinite dissipation, $\overline{\sigma}\to\infty$, for which 
$T_{\rm cl} \to 0$. The crossover scale we obtain is similar in form
to the estimate, $T_{\rm cl} \approx T_c/\overline{\sigma}$, given in 
ref.~\cite{carlson99}, but is much larger in magnitude.

\subsection{Low temperature behavior}

We next turn to the temperature dependence of the renormalized
stiffness in the presence of dissipation, where we have set the bare stiffness 
to be independent of temperature. This is of course an unphysical
assumption for a $d$-wave SC, but our aim is to explicitly check whether
a linear $T$-dependence can be obtained within a model of purely 
dissipative phase fluctuations even when temperatures are smaller than the 
thermal crossover scale estimated above. 
The fluctuation $\la\delta\theta^2_{_\pll}\ra$ at low $T$ 
can be evaluated analytically again by setting the cutoff $n_c \to \infty$.
We can then cast the Matsubara sum in the form
\be
\sum_{n=-\infty}^{\infty} \frac{A(\vQ,T)}{n^2+ B(\vQ,T)|n|+
C(\vQ,T)}
=2 \sum_{n=0}^{\infty}\frac{A(\vQ,T)}{n^2+ B(\vQ,T) n+
C(\vQ,T)} -\frac{A(\vQ,T)}{C(\vQ,T)}. \nonumber
\ee
Rewriting the denominator of the first term in the form
$(n+n_1) (n+n_2)$, we separate out the terms using partial fractions and 
express the resulting sums in terms of digamma functions. As $T\to
0$, $n_{1,2} \to \infty$ which allows us to use the asymptotic expansion for
the digamma function. The linear $T$ term arising from the
infinite sum is precisely canceled by the linear $T$ term from
the $A(\vQ,T)/C(\vQ,T)$ term, leaving only a quadratic
temperature dependence as was pointed out in ref.~\cite{millis98}. 
We thus finally arrive at $\la\delta\theta^2_{_\pll}\ra(T)=
\la\delta\theta^2_{_\pll}\ra(0) +(\overline{\sigma}/3) (T/D_{_\pll}d_c)^2$
at low $T$, from which
\be
{{D_{_\pll}(T)}\over{D_{_\pll}(0)}} \approx 1-\frac{\overline{\sigma}}{6}
\left(\frac{T}{D_{_\pll}(0)d_c}\right)^2.
\label{asymptote}
\ee
Thus, ignoring the effects of nodal quasiparticles, 
the asymptotic low temperature stiffness decreases as $T^2$
in the presence of dissipation. 

At high temperature, above the thermal crossover scale, we recover the 
classical result $\la\delta\theta^2_{_\pll}\ra(T) \approx 2 T /(D_{_\pll}d_c)$.
An improved estimate
of the thermal crossover scale $T_{\rm cl}$ is obtained by matching the 
slope of this high temperature result for $\la\delta\theta^2_{_\pll}\ra$ 
with the low temperature result of Eq.~(\ref{asymptote}). This gives us 
$T_{\rm cl}=3 D_{_\pll} d_c/\overline{\sigma}$ as stated earlier.

At this stage, we turn to the recent results of Lemberger and
co-workers \cite{lemberger00.theory}, who use a circuit analogy and
model a Josephson junction as an inductance ($L_0$) shunted
by a resistance ($R$) and capacitance ($C$). To make correspondence with
this work, we note that the inductance $L_0 \sim (D_{_\pll}d_c)^{-1}$, 
the charging energy
$e^2/2C \sim (e^2/\epsilon_b \xi_0)$ and the resistance $R \sim 
(1/\overline{\sigma})(h/e^2) $. Upto numerical factors of order unity,
our expressions 
for the magnitude of quantum fluctuations and the thermal crossover scale 
are then in agreement. The predicted quantum to thermal crossover has also
been recently observed in experiments
on conventional s-wave superconducting films\cite{lemberger00.expt}.

\subsection{Numerical Results}

In order to obtain the various scales for the cuprates, we will choose 
parameters of the above action appropriate for the bilayer system $Bi2212$
and evaluate the above estimates. We then present results of detailed 
numerical calculations which are shown to agree with these simple estimates.

In the absence of detailed information on the bilayer couplings, we make
the assumption that the two layers within a bilayer are strongly coupled
and phase locked.  Experimentally, the in-plane penetration depth of
optimally doped $Bi2212$ is around $2100\AA$ and this translates into a
bilayer stiffness $\approx 75 meV$.  We use $\epsilon_\infty \approx 10$,
and $d_c/a \approx 4$, with the in-plane coherence length $\xi_0/a \approx
10$. This leads to a Coulomb scale $(e^2/\epsilon_\infty\xi_0)\approx 35
meV$.  Using the above parameters, we find large quantum fluctuations in
the non-dissipative case ($\overline{\sigma}=0$)  with
$\la\delta\theta^{2}_{_\pll}\ra \gtrsim 1$. The thermal crossover scale as
estimated from the zero temperature bilayer stiffness, $T_{\rm cl} \gg T_c
\sim 100 K$. A more sensible estimate is obtained by considering the
temperature dependence of the bare stiffness, and this leads to a
crossover scale for $T_{\rm cl} \sim T_c$ for $\overline{\sigma}=0$. 

To study the effect of dissipation, we use conductivity data obtained from
experiments performed in the superconducting
state. Consistent with our assumption of strongly coupled phases within a
bilayer for $Bi2212$, the dissipation parameter $\overline{\sigma}$ for
this system will be taken to be the dimensionless {\it bilayer} conductivity.
Recent measurements by Corson {\it et al}\cite{corson00} on $Bi2212$ films
give a Drude conductivity with a large low frequency value corresponding 
to $\overline{\sigma}\approx 75$ and a width of a few
Terahertz. Similar large conductivities have been measured in the microwave 
regime
\cite{sflee96}. We note that the ``universal'' quasiparticle conductivity
predicted by Lee \cite{palee93} for the bilayer conductivity (at
$T=0,\omega\to 0$) corresponds to
\cite{footnote.sigma} $\overline{\sigma}\approx 24$ for $Bi2212$.
The difference between this ``universal'' value, and the 
$\overline{\sigma}\approx 150$ inferred from microwave data
\cite{sflee96} may be due to vertex corrections\cite{durst99}.
In our calculations, we use a {\em constant} dissipation 
with $\overline{\sigma}\approx 150$ (as an {\em overestimate}) over the 
entire frequency range of interest: $\omega_n$ with $|n| \leq n_c$. This
frequency range corresponds to $\omega\lesssim 100 meV$ at $T=0$.

In the presence
of dissipation, we find that quantum fluctuations are reduced to a very
small value $\la\delta\theta^2_{_\pll}\ra\lesssim 0.2$. The thermal
crossover scale is then $T_{\rm cl}=3 D_{_\pll} d_c/\overline{\sigma}\sim 
18 K$. This is consistent with our numerics, where
we find that linear $T$ behavior from thermal phase fluctuations only sets
in above a temperature $\sim 20 K$, for this magnitude of dissipation.
While this is a low temperature scale, penetration depth measurements
\cite{jacobs95,sflee96} observe
a smooth linear $T$ behavior down to much lower temperatures $\sim 5 K$,
which cannot be reconciled with this crossover scale.

Turning to $YBCO$, and treating this system as weakly coupled single layers, 
far infra-red reflectance measurements
\cite{basov98} appear to be consistent with $\overline{\sigma}\sim
10$-$15$ over a wide frequency range $\sim 5$-$100 meV$. This smaller
value of $\overline{\sigma}$ compared to $Bi2212$ implies that dissipative 
effects are less important in $YBCO$.
The ``universal'' conductivity \cite{palee93} for this case
\cite{footnote.sigma} corresponds to $\overline{\sigma}\approx 9$, not
inconsistent with the above data.
However, low frequency microwave measurements\cite{hosseini99} on $YBCO$
observe a strong frequency and temperature dependent quasiparticle 
conductivity, of the Drude form. We have checked that using a temperature 
dependent Drude conductivity in this very low frequency regime, in addition 
to a constant dissipation over the entire frequency range, does not 
significantly affect our results.

\begin{figure}[h]
\begin{center}
\includegraphics[width=8cm, angle=0]{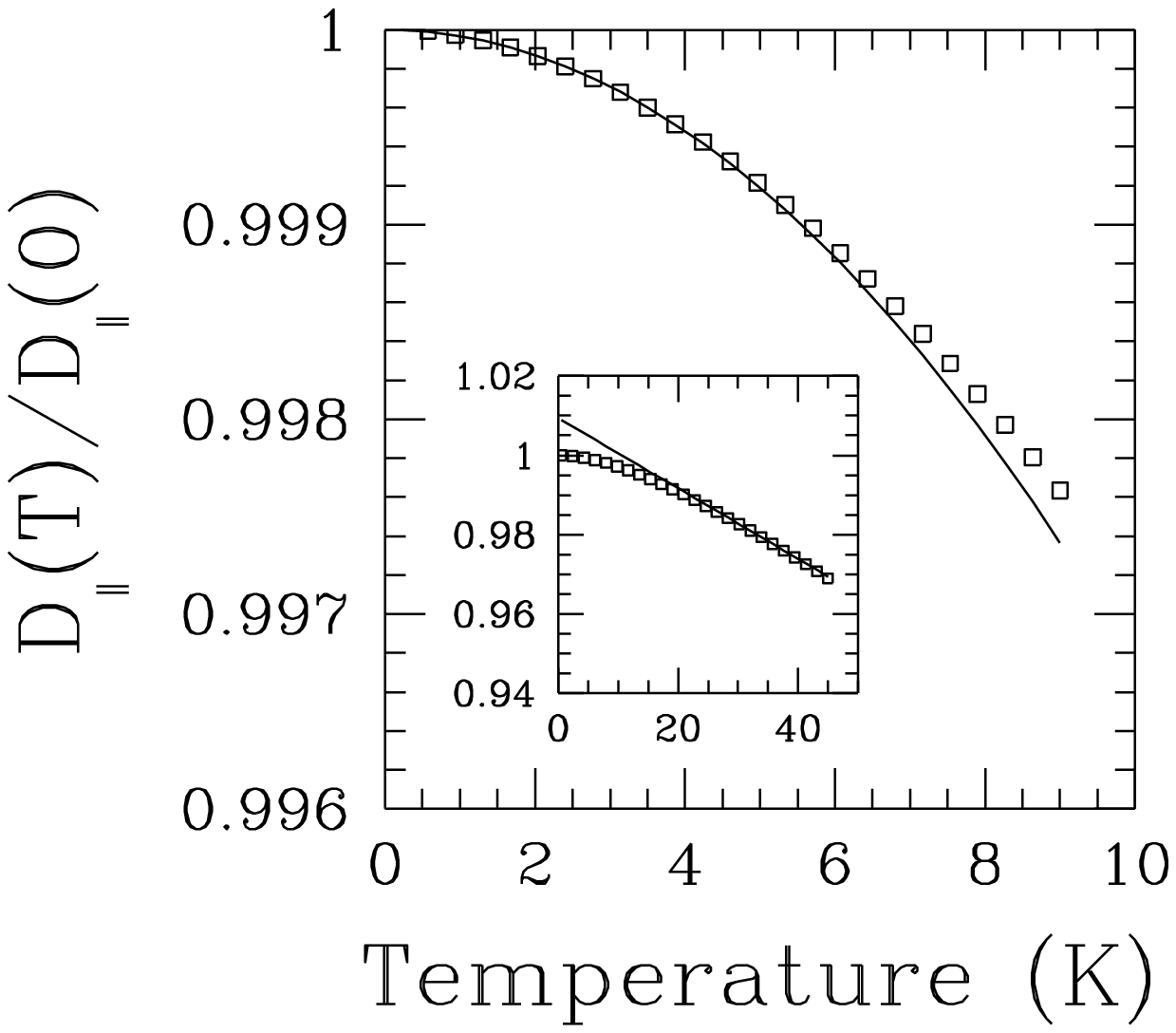}
\end{center}
\caption{\footnotesize Low temperature behavior of the renormalized 
superfluid stiffness for the dimensionless dissipation
$\overline{\sigma}=150$. The bare stiffness was chosen to be
$D^0_{_\pll}d_c=80 meV$, independent of temperature, and the renormalized 
$D_{_\pll}d_c (T=0) \approx 75
meV$, corresponding to $\la\delta\theta_{_\pll}^2 (T=0)\ra\approx 0.1$. The
values obtained from the numerics are shown as squares while the solid
line is the analytical $T^2$ form given in Eq.~(\ref{asymptote}). Linear
temperature dependence sets in above a temperature $\sim 20 K$
as seen from the asymptote in the inset, which compares well with the 
estimate of the thermal crossover scale $3 D_{_\pll}d_c/\overline{\sigma} 
\approx 18 K$. 
}
\label{fig1}
\end{figure}

In Fig.~\ref{fig1} we compare our analytical results and estimates 
obtained above, for $Bi2212$, with a numerical 
solution of the SCHA equations (\ref{scha}) and (\ref{flucn}).
We find that the purely quadratic dependence persists up to about
$6 K$ for the parameters discussed above, while linear T dependence
only sets in at temperatures $\gtrsim 20 K$, consistent with the above
estimate for $T_{\rm cl}$. Thus, it is impossible
to ignore quasiparticles for understanding the smooth linear $T$
behavior of the penetration depth which has been
observed\cite{jacobs95,sflee96} down to temperatures $\sim 5 K$. 

\begin{figure}[h]
\begin{center}
\includegraphics[width=8cm, angle=0]{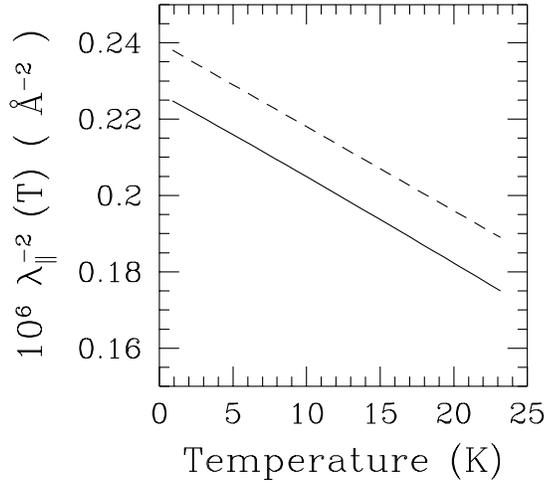}
\end{center}
\caption{\footnotesize The bare and renormalized $1/\lambda^2_\pll(T)$ 
plotted using dashed and solid line respectively, for a dimensionless
dissipation $\overline{\sigma}=150$. The bare value of
$\lambda_{\pll,0}$ and its slope were chosen such that the renormalized
values, calculated using the SCHA equations (\ref{scha}) and
(\ref{flucn}), $\lambda_\pll\approx 2100 \AA$ and its slope $d\lambda_\pll/dT
\approx 10.0 \AA/K$, are in agreement with experiments \protect\cite{sflee96}
on $Bi2212$. The renormalization due
to quantum fluctuations is seen to be $\approx 5\%$, much smaller than
the $\approx 50\%$ renormalization obtained in the non-dissipative case
\protect\cite{arun00}.}
\label{fig2}
\end{figure}

We next include the linear $T$ effect of quasiparticle
excitations in the bare stiffness and ask how dissipative
phase fluctuations renormalize this. The numerical result is plotted in 
Fig.~\ref{fig2} and shows that both the $T=0$ stiffness and its slope are 
renormalized by small amounts. This is completely consistent with our 
estimates for a small quantum renormalization in the presence of dissipation
and a temperature scale of about $20 K$ below which classical thermal effects 
are unimportant.

\subsection{High temperature behavior}

Although quasiparticles dominate at low temperature, eventually
phase fluctuations do become important at higher temperatures, in driving 
the transition to the non-superconducting state.
The approximation used thus far (SCHA) by itself is clearly
inadequate to address the problem of $T_c$ since it only includes
longitudinal phase fluctuations. We thus proceed in two steps:
first we calculate the temperature dependence of the superfluid stiffness
within the SCHA, and then we supplement it with the Nelson-Kosterlitz
condition for the universal jump in the stiffness at a
2D Kosterlitz-Thouless transition \cite{chaikin} 
mediated by the unbinding of vortices. We do not take into account,
for simplicity, the effect of layering in this calculation, but this could
be easily done using well-known results for dimensional crossover. 

\begin{figure}[h]
\begin{center}
\includegraphics[width=8cm, angle=0]{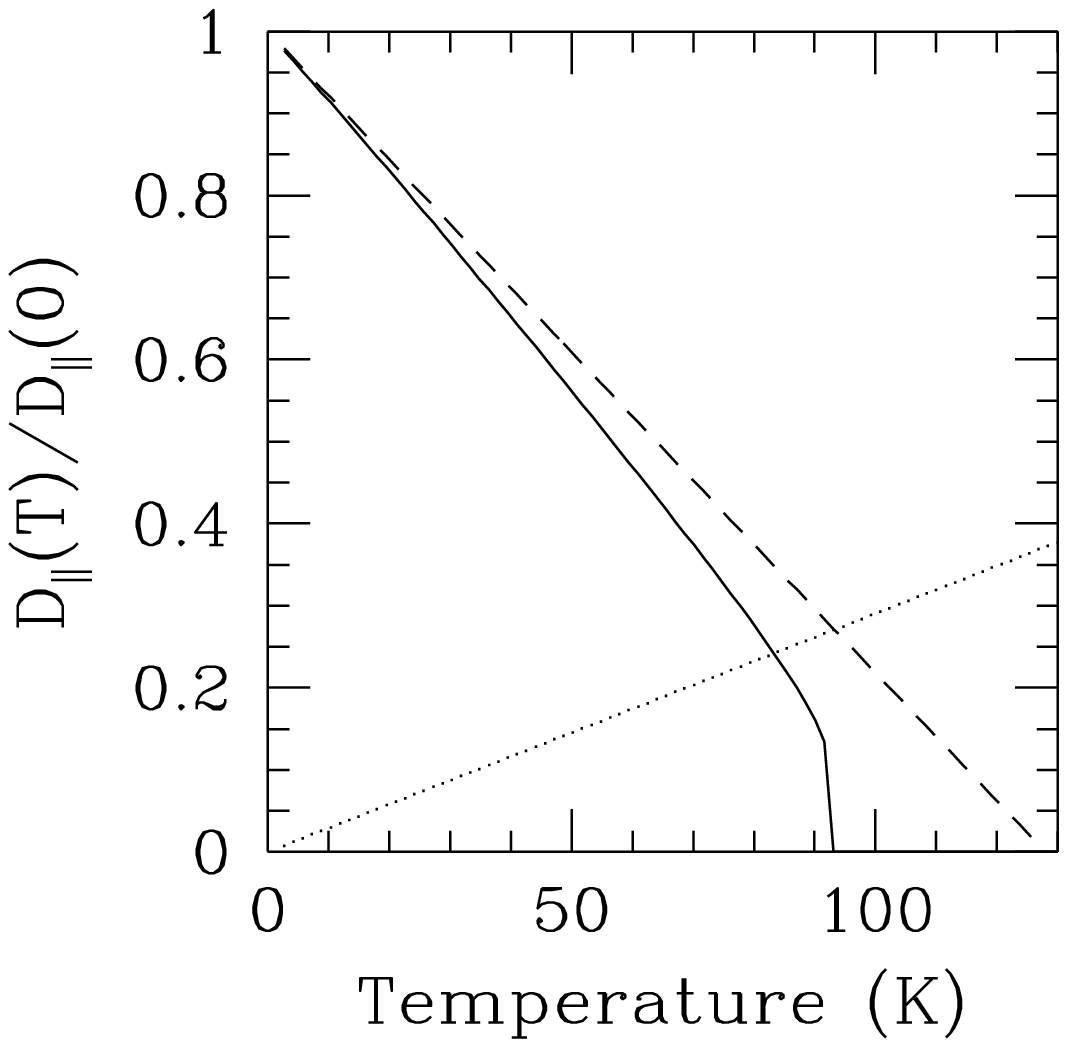}
\end{center}
\caption{\footnotesize The bare and renormalized stiffness plotted 
as a function of temperature, using dashed and solid lines respectively, 
to obtain an estimate of the transition temperature $T_c$ in $Bi2212$. 
We chose the bare bilayer stiffness $d_c D^0_{_\pll}(T)= d_c D^0_{_\pll}(0)- 
\alpha^0 T$ with parameters
$d_c D^0_{_\pll}(0)\approx 80 meV$ and $\alpha^0\approx 
0.7 meV/K$, relevant to bilayer $Bi2212$.
This leads to a renormalized bilayer stiffness of $\approx 75 meV$ and a low
temperature slope $\approx 0.7 meV / K$ for the bilayer stiffness,
consistent with low temperature penetration depth experiments
\protect\cite{sflee96} in $Bi2212$.
The renormalized stiffness is computed using the SCHA equations
(\ref{scha}) and (\ref{flucn}) with a dimensionless dissipation
$\overline{\sigma}=150$. The renormalized stiffness (solid line), within 
the SCHA,
shows a jump near $T\sim 90 K$, but that is likely an artifact of the
approximation. The temperature at which {\em transverse
excitations} drive the transition to a non-superconducting state is
estimated from the Nelson-Kosterlitz condition, as the point at which 
the dotted line intersects the
{\em renormalized stiffness} curve (see text for details). This temperature,
$T_{KT} \approx 80 K$, is in reasonable agreement with experimental
values\protect\cite{sflee96} for $T_c\sim 90 K$ in optimal $Bi2212$.}
\label{fig3}
\end{figure}

Our numerical results, obtained by solving the SCHA equations for
parameter values relevant to $Bi2212$, are plotted in Fig.~\ref{fig3}.  
We take the bare stiffness (the dashed line in Fig.~\ref{fig3}) to
decrease linearly with temperature, consistent with experiments, as a
result of quasiparticle excitations. We extend this linear $T$ dependence
of the bare stiffness to higher temperatures assuming that the gap
function is unaffected for $T < T_c$. This is probably a good assumption
in underdoped systems and may not be unreasonable for optimal doping. The
renormalized stiffness calculated within SCHA is shown as the full line.
This stiffness shows a jump at a temperature $\sim 90 K$, but that is 
likely an artifact of the SCHA.
The Kosterlitz-Thouless transition occurs at $T_{KT} = \pi D_s(T_{KT}^-)
d_c/8$, where the {\em fully} renormalized stiffness $D_s$ is evaluated
just below $T_{KT}$. We use the renormalized stiffness within the SCHA and
the above condition, to obtain an approximate location of this transition.  
As can be seen from Fig.~\ref{fig3}, this gives us a reasonable estimate
of $T_c\approx 80 K $, when compared with
experiments\cite{jacobs95,sflee96} on optimal $Bi2212$ which give 
$T_c \approx 90 K$.


\section{conclusions}

In this paper we have derived and analyzed the effective action for phase
fluctuations in the presence of dissipation arising from low-frequency
absorption in a $d$-wave superconductor. We find that including
dissipation reduces the magnitude of quantum phase fluctuations. The
temperature at which one crosses over to thermal phase fluctuation is also
reduced drastically. However, for parameter values relevant to the
high-$T_c$ cuprates, we find that the thermal crossover scale is still
large, so that quasiparticles dominate the asymptotic low temperature
properties. In particular, they must be responsible for linear
$T$-dependence of the low temperature penetration depth.

\bigskip

{\bf Acknowledgements:} We acknowledge C. Di Castro, M. Grilli,
S. de Palo and T.V. Ramakrishnan for useful discussions and
comments. We are particularly grateful to T. Lemberger for helpful
discussions and detailed comments on the paper. The work of M.R. was 
supported in part by the D.S.T., Govt. of India, under the Swarnajayanti 
scheme. 

\bigskip
\vfill\eject

\appendix
\section{Two fluid hydrodynamics and collective modes at low
temperature}

In this Appendix we analyze the finite-temperature properties of the 
two-fluid model, to determine the collective modes of a superconductor 
in the presence of dissipative processes. We first consider a 3D Galilean
invariant system and later generalize the result to a layered system,
still maintaining Galilean invariance in the planes. Our goal is
to determine the phase action that gives rise to this collective mode,
which serves as another way to arrive at the Gaussian action derived
microscopically in the text.

The ordinary equations of a superfluid \cite{Landau.Fluid} 
are altered by the addition of a dissipative contribution. 
The longitudinal modes arising in the presence of
long-range Coulomb forces obey the following set of 
linearized equations:
\begin{eqnarray}
\label{eqsf1}
& &\omega\delta\rho=\vq\cdot\vj,\\
\label{eqsf2}
& &\omega\delta s+\frac{s\rho_s}{\rho}\vq\cdot(\vs-\vn)=0,\\
\label{eqsf3}
& &\vj=\left[{\rm i} \frac{e \rho_s}{m\omega}+
\left({{\sigma_{reg}(\omega)}\over{e}}\right) \right] 
\vE + \frac{\vq}{\omega}\delta P,\\
\label{eqsf4}
& &{\rm i}\omega \vs-{\rm i} \vq \delta\mu+\frac{e\vE}{m}=0,\\
\label{eqsf5}
& &{\rm i}\vq\cdot\vE=\frac{4\pi e}{\epsilon_\infty}\delta\rho.
\end{eqnarray}
Here the symbols have their usual meanings:
$\omega,\vq$ are the frequency and wavevector,
the subscripts $s,n$ refer to the superfluid or normal component,
$\rho,\vj$ are the particle density and particle-current density,
$\vv$ indicates a velocity, $\vE$ 
is the internal longitudinal 
electric field, $\sigma_{reg}(\omega)$ is the regular part of
the complex conductivity, and
$\epsilon_\infty$ is the background dielectric constant.
Further the thermodynamic variables, the pressure $P$,
the entropy per particle $s$, and the chemical potential $\mu$, 
are related by the identity
\be
\delta\mu=-s\delta T+\frac{1}{\rho}\delta P.
\ee

Note that dissipative processes due to thermal conductivity, 
which should appear on the right-hand side of Eq.~(\ref{eqsf3}), 
can be neglected at low temperature and anyway, affect only second 
sound, which is decoupled from the density mode, as we shall see below. 
Moreover, the Lorentz force and the Joule heating are second order 
in the fluctuations, and do not affect the 
linearized equations which we are investigating. 

The equations for first and second sound, i.e., for density and 
entropy fluctuations respectively, are in principle coupled via the pressure 
variations:
\bea
\omega^2 \delta\rho&=& \left[ \frac{4\pi e^2 \rho_s}{m\epsilon_\infty}-
{\rm i}\frac{4\pi\omega \sigma_{reg}(\omega)}{\epsilon_\infty}\right] \delta\rho+q^2 
\delta P,\\ \nonumber
\omega^2\delta s&=&q^2 \frac{\rho_s s^2}{\rho_n} \delta T-
\left[\frac{4\pi e^2 \rho_n}{m\epsilon_\infty}+ 
{\rm i}\frac{4\pi \omega \sigma_{reg}(\omega)}{\epsilon_\infty}\right]
\frac{\rho_s s} {\rho_n \rho} \delta\rho.
\label{eqlmr}
\eea
However, observing that we can rewrite
\begin{equation}
\delta P= \left(\frac{\partial P}{\partial \rho}\right)_T \delta\rho+
\left(\frac{\partial P}{\partial T}\right)_{\rho} \delta T,
\label{eqdp}
\end{equation}
and using $\left(\partial P / \partial T\right)_{\rho} = 0$ as
$T \to 0$ as a consequence of Nernst's theorem \cite{Landau.SP1},
we can conclude that, at low temperature, second sound does not mix with
the longitudinal density modes.
From equations (\ref{eqlmr}) and (\ref{eqdp}), we then
deduce the the dispersion for density fluctuations, given by
\begin{equation}
\omega^2 = \left[ \frac{4\pi e^2 \rho_s}{m\epsilon_\infty}- 
{\rm i}\frac{4\pi \omega \sigma_{reg}(\omega)}{\epsilon_\infty}\right]+ c_p^2 q^2 
\label{rhomode}
\end{equation}
where $c_p$ is a constant. From now on we neglect the term $c_p^2 q^2$
which is unimportant in the long wavelength limit.
This is the plasmon dispersion, and coincides with
result of the microscopic derivation in ref.~\cite{arun00}.
(To make connection with Eq.~(29) of that reference, note that
the real part of the right hand side of Eq.~(\ref{rhomode}) above
is $4\pi\omega {\rm Im}\sigma(\omega)$, where $\sigma(\omega)$ is the 
{\it total} conductivity.)

Since we are interested in layered systems
Eq.~(\ref{rhomode}) must be modified.
For simplicity we consider the case of
carriers confined to stacked (Galilean invariant) planes 
interacting via an the anisotropic Coulomb potential $V(\vq)$
defined in Eq.~(\ref{coulomb}). Referring to the components in-plane and
across planes with a subscript $\parallel$ and $\perp$ respectively, we then
assume that $m_\perp=\infty$, $m_\parallel=m$, $\sigma_\perp=0$ and
denoting $\sigma^{reg}_\parallel = \sigma_\parallel$.
The longitudinal electric field $E$ 
has components, $E_\parallel=-{\rm i} q_\parallel \phi(\vq)$ and 
$E_\perp=-{\rm i} q_\perp \phi(\vq)$, where the
the electrostatic potential $\phi(\vq)$ for a
density disturbance $\delta\rho(-\vq)$ is given by
$\phi(\vq)=(V(\vq)/e) \delta\rho$. 
As $m_\perp=\infty$ and $\sigma_\perp=0$, only the in-plane component 
$E_\parallel$ enters in the linear response response Eq.~(\ref{eqsf3}) and 
determines the ballistic motion of the superfluid electrons, 
Eq.~(\ref{eqsf4}). Rewriting (\ref{eqsf1})-(\ref{eqsf5}) and decoupling again 
the first and second sound as before, we obtain the density mode 
\be
\omega^2 = \left[ \frac{\rho_s}{m}- 
\frac{{\rm i}\omega \sigma_\parallel(\omega)}{e^2}
\right] q_\parallel^2 V(\vq). 
\label{rhoan}
\ee
Eq.~(\ref{rhoan}) allows us to deduce the correct expression for the
phase-only Lagrangian in the layered case. Indeed, according to the previous
equation, in the presence of dissipation, the superfluid stiffness
must be transformed as $\rho_s/m \to \left(\rho_s/m -
{\rm i}\omega\sigma_\parallel(\omega)/e^2\right)$.
We thus obtain, at the Gaussian level, the Lagrangian density
\be
{\cal{L}} ({\bf q}, \omega)=
\frac{1}{8} \left[ V^{-1}(\vq) \omega^2 
-\left(\frac{\rho_s}{m}
-\frac{ {\rm i}\omega\sigma_\parallel(\omega)}{e^2}
\right) q_{_\pll}^2 \right] |\theta(\vq, \omega)|^2,
\ee
which is the same as the Gaussian action used in the paper.


\section{Analytical results for phase fluctuations with arbitrary 
$\overline{\sigma}$}

It is useful to calculate the
the fluctuation $\la\delta\theta^2_{_\pll}\ra$ for {\em arbitrary} 
dissipation $\overline{\sigma}$, even approximately.
To make progress we work with the anisotropic
Coulomb interaction 
\be
V(\vq) = \frac{4\pi e^2}{\varepsilon_{_\pll} q^2_{_\pll} + 
\varepsilon_{_\perp} q^2_{_\perp}}.
\label{aniso_coul}
\ee
which is different from the Coulomb potential (\ref{coulomb})
for layered systems used in the paper. In particular, one cannot take
a 2D limit of (\ref{aniso_coul}) unlike in the layered case. However,
it permits us a simple evaluation of the integrals appearing for
$\la\delta\theta^2_{_\pll}\ra$ in (\ref{flucn}).

On using the appropriate scaled Coulomb interaction and simplifying
$\gamma_{_\pll}(\vQ)\simeq Q_{_\pll}^2$, the renormalized Gaussian action 
(with $D_{_\perp}=0$) takes the form
\be
S_{G}[\theta] = \frac{1}{8 T}{\sum_{\vQ,\omega_n}}
\left[\frac{\varepsilon_{_\pll} \omega^2_n d_c}
{4\pi e^2}\left( 1 + \eta \frac{Q_{_\perp}^2}{
Q_{_\pll}^2}\right)
+ \left( D_{_\pll} d_c + \frac{\overline{\sigma}}{2\pi}|\omega_n|\right)
\right] Q_{_\pll}^2 |\theta(\vQ,n)|^2
\label{simple}
\ee
with $\eta\equiv(\varepsilon_{_\perp}\xi_0^2)/(\varepsilon_{_\pll} d_c^2)$.
Setting $\omega_c \equiv 4\pi e^2/d_c$, we then obtain the 
fluctuation
\be
\la\delta\theta^2_{_\pll}\ra
= 2 T \omega_c
\int_{-\pi}^{\pi} \frac{dQ_{_\perp}d^2 Q_{_\pll}}{(2\pi)^3}
\sum_{\omega_n} \left[
\varepsilon_{_\pll}(1+\eta \zeta_{\vQ})\omega_n^2+ \omega_c
\left(D_{_\pll}d_c+\frac{\overline{\sigma}}{2\pi} |\omega_n|\right)
\right]^{-1},
\label{flucn.1}
\end{equation}
with $\zeta_{\vQ}\equiv Q_{_\perp}^2/Q_{_\pll}^2$.

The integrand in (\ref{flucn.1}) depends on $\vQ$ only through 
$\zeta_{\vQ}$. The ${\vQ}$ integral can therefore be transformed to an
integral over the variable $\zeta$, with density
$N(\zeta)=1/3\sqrt\zeta$ for
$\zeta\leq 1$ and $N(\zeta)=1/3\zeta^2$ for $\zeta>1$, and
\begin{equation}
\la\delta\theta^2_{_\pll}\ra=\int_0^\infty  d\zeta N(\zeta)
\Phi((1+\eta\zeta)\varepsilon_\parallel) \equiv
\Phi({\bar \varepsilon}) \int_0^\infty  d\zeta N(\zeta)=
\Phi({\bar \varepsilon}),
\label{eqden}
\end{equation}
where
\begin{equation}
\Phi(\varepsilon)=2 T \omega_c \sum_{\omega_n} \left[
\varepsilon\omega_n^2+\omega_c \left(D_{_\pll}d_c +\frac{\overline{\sigma}}
{2\pi} |\omega_n|\right)\right]^{-1},
\label{Phi}
\end{equation}
and $\bar \varepsilon$ is a suitable average value of $(1+\eta\zeta)
\varepsilon_{_\pll}$. We have written (\ref{eqden}) such that the effect
of the $\vQ$ integral appears as a renormalization of the bare in-plane 
dielectric
constant $\varepsilon_\parallel$ to a larger value $\bar \varepsilon$.
Of course, the value of $\bar \varepsilon$ is temperature dependent.
However, the leading temperature dependence of
$\la\delta\theta^2_{_\pll}\ra$ 
is, in most cases, $\overline{\varepsilon}$-independent 
(with the noticeable exception of the dissipation-less case 
$\overline{\sigma}=0$). Therefore, to proceed further analytically, we take
$\overline{\varepsilon}$ to be a constant henceforth.

The quantum corrections can be calculated analytically, expressing
$\la\delta\theta^2_{_\pll}\ra$ by means of the spectral representation for the
Matsubara
phase propagator deduced from Eq. (\ref{Phi}), so that
\begin{equation}
\la\delta\theta^2_{_\pll}\ra(T)=\frac{2 \omega_c}{\overline{\varepsilon}}
\int_{-\infty}^{+\infty} \frac{dz}{2\pi} \frac{z\omega_c\overline{\sigma}/
2\pi\overline{\varepsilon}}
{(z^2-\omega_p^2)^2 +z^2(\overline{\sigma}\omega_c/2\pi\overline{\varepsilon}
)^2}\coth\left(\frac{z}{2 T}\right),
\label{eq8}
\end{equation}
where $\omega_p^2=(D_{_\pll}d_c)~(\omega_c/\overline{\varepsilon})=
4\pi e^2\rho_s/\overline{\varepsilon} m^*$.

The above expression allows us to
explicitly calculate the value of $\la\delta\theta^2\ra(T=0)$.
Introducing the dimensionless parameter 
$s=(\overline{\sigma}/4\pi)
~(\omega_c/\overline{\varepsilon}\omega_p)=(\overline{\sigma}/4\pi)~
\sqrt{\omega_c/\overline{\varepsilon}D_\parallel d_c}$,
and evaluating the above integral at zero temperature we obtain the
zero-point-motion contribution
\be
\la\delta\theta^2_{_\pll}\ra(0)=
\frac{2 \omega_c}{2\pi \overline{\varepsilon}\omega_p\sqrt{s^2-1}}
\ln \left(\frac{s+\sqrt{s^2-1}}{s-\sqrt{s^2-1}} \right)
\label{eq9}
\end{equation}
in the case $s>1$, which is physically relevant for the cuprates, and
\begin{equation}
\la\delta\theta^2_{_\pll}\ra(0)=
\frac{2 \omega_c}{\pi\overline{\varepsilon}\omega_p\sqrt {1-s^2}}
\arctan{\frac{\sqrt{1-s^2}}{s}}
\label{eq9a}
\end{equation}
in the case $s<1$. For $s\to 1$ both the above expressions (\ref{eq9}) and
(\ref{eq9a}) reduce to ${2 \omega_c/\pi \overline{\varepsilon}\omega_p}$.
For $s  \gg 1$, from Eq. (\ref{eq9}), $\la\delta\theta^2_{_\pll}\ra(0)
\simeq (8/\overline{\sigma})~\ln(\frac{\overline{\sigma}}{2\pi}
\sqrt{\omega_c/(\overline{\varepsilon}D_{_\pll}d_c)})$ in agreement with 
the expression
used in the text, except the relevant energy scale, $\omega_c=4\pi e^2/d_c$, 
for the anisotropic Coulomb potential appears instead of 
$2\pi e^2/\xi_0$, the scale appropriate when large $Q_{_\pll}$
contributes and we are closer to a 2D limit in the fluctuation integral.
The quantum fluctuations are large for small $s$ and decrease monotonically 
with increasing $s$. 
For $s\to \infty$, $\la\delta\theta^2_{_\pll}\ra \to 0$, and the
classical limit is recovered. 

At finite temperature and for a bare stiffness which is independent of 
temperature, the analytical result for the quadratic 
temperature dependence of $\la\delta\theta^2_{_\pll}\ra(T)$ for arbitrary 
$\overline{\sigma}$, given in Eq.(\ref{asymptote}), has been derived in 
the text. The same result also follows from a low temperature analysis 
of Eq.(\ref{eq8}) above.


\vfill\eject

\vfill\eject


\begin{thebibliography}{999}
\bibitem{footnote.nonlocal}
As we shall see, the damping of quantum fluctuations at low temperature 
is not determined by nodal quasiparticles, but rather by all electronic
excitations in the range $\omega \lesssim 2 \Delta$. The mean free path 
of these excitations (estimated from the measured conductivity) is 
smaller than both the penetration depth and the skin depth, thus 
validating the local approximation.
\bibitem{footnote1}
As shown in ref.~\cite{arun00} the scaled interaction in the action (\ref{xy1})
is $\tilde{V}(\vQ)\equiv V(\vQ_{_\pll}/\xi_0,Q_{_\perp}/d_c)$.
The following plausibility argument may help one to see why this is so.
In the 2D limit ($d_c \to \infty$) one finds that
$\tilde{V}(\vQ)/(\xi_0^2 d_c) \to 2\pi e^2/(\epsilon_\infty Q_{_\pll}\xi_0)$.
This leads to an energy scale describing interactions on the coarse-grained
lattice, with characteristic $Q_{_\pll} \sim 1$, of the order of
$e^2 /(\epsilon_\infty \xi_0)$, as expected on physical grounds.
\bibitem{footnote.sigma}
The ``universal conductivity'' for a single layer is given by $\sigma_{2D}=
(e^2/h)~(2/\pi)~(v_{_F}/v_{2})$ where the ratio $(v_{_F}/v_{2})\approx 20$
from ARPES measurements \cite{mesot99,chiao99}.
\end{thebibliography}
\end{document}